\documentclass[prb,amsmath,amsfonts,amssymb,twocolumn,floatfix,unsortedaddress]{revtex4-1}
\usepackage{graphicx}
\usepackage{bm}
\usepackage{color}

\begin{document}
\title{Polynomial Similarity Transformation Theory: A smooth interpolation between coupled cluster doubles and projected BCS applied to the reduced BCS Hamiltonian}
\author{Matthias Degroote}
\affiliation{Department of Chemistry, Rice University, Houston, TX 77005-1892}

\author{Thomas M. Henderson}
\affiliation{Department of Chemistry and Department of Physics and Astronomy, Rice University, Houston, TX 77005-1892}

\author{Jinmo Zhao}
\affiliation{Department of Chemistry, Rice University, Houston, TX 77005-1892}

\author{Jorge Dukelsky}
\affiliation{Instituto de Estructura de la Materia, CSIC, Serrano 123, E-28006 Madrid, Spain}

\author{Gustavo E. Scuseria}
\affiliation{Department of Chemistry and Department of Physics and Astronomy, Rice University, Houston, TX 77005-1892}
\date{\today}

\begin{abstract}
We present a similarity transformation theory based on a polynomial form of a particle-hole pair excitation operator. In the weakly correlated limit, this polynomial becomes an exponential, leading to coupled cluster doubles.  In the opposite strongly correlated limit, the polynomial becomes an extended Bessel expansion and yields the projected BCS wavefunction.  In between, we interpolate using a single parameter. The effective Hamiltonian is non-hermitian and this Polynomial Similarity Transformation Theory follows the philosophy of traditional coupled cluster, left projecting the transformed Hamiltonian onto subspaces of the Hilbert space in which the wave function variance is forced to be zero.  Similarly, the interpolation parameter is obtained through minimizing the next residual in the projective hierarchy.  We rationalize and demonstrate how and why coupled cluster doubles is ill suited to the strongly correlated limit whereas the Bessel expansion remains well behaved.  The model provides accurate wave functions with energy errors that in its best variant are smaller than 1\% across all interaction stengths.  The numerical cost is polynomial in system size and the theory can be straightforwardly applied to any realistic Hamiltonian. 
\end{abstract}
\maketitle

\section{Introduction}
Simple model Hamiltonians are important primarily because their solutions provide guidelines which help elucidate the physics of more complicated realistic Hamiltonians.  This is particularly true if the model Hamiltonian can be solved exactly.  One such model Hamiltonian is the pairing or reduced BCS Hamiltonian, which takes the simple form
\begin{subequations}
\begin{align}
H &= \sum_p \epsilon_p \, N_p - G \, \sum_{pq} P_p^\dagger \, P_q,
\\
N_p &= c_{p_\uparrow}^\dagger \, c_{p_\uparrow} + c_{p_\downarrow}^\dagger \, c_{p_\downarrow},
\\
P_p^\dagger &= c_{p_\uparrow}^\dagger \, c_{p_\downarrow}^\dagger,
\\
P_p &= c_{p_\downarrow} \, c_{p_\uparrow},
\end{align}
\end{subequations}
where $p$ and $q$ index single-particle levels.  This Hamiltonian phenomenologically describes the Cooper problem of bound electron pairs, created by the pair creation operators $P_p^\dagger$, interacting attractively with the holes they have left behind in the Fermi sea, and its importance lies in that it constitutes a simple model for superconductivity.  Note that seniority is a symmetry of the Hamiltonian, where the seniority of a determinant is the number of singly-occupied single-particle levels, so that its eigenstates can be labeled by their seniorities.  

Importantly for our purposes, this Hamiltonian can be solved exactly.\cite{Richardson1963,Richardson1966}  The ground state wave function for $n_p$ pairs of particles occupies $n_p$ distinct geminals (two-particle states):
\begin{subequations}
\begin{align}
|\Psi\rangle &= \prod\limits_{\mu=1}^{n_p} \Gamma_\mu^\dagger |-\rangle,
\\
\Gamma_\mu^\dagger &= \sum_p \frac{1}{R_\mu - \epsilon_p} P_p^\dagger,
\end{align}
\end{subequations}
where $|-\rangle$ is the physical vacuum and the parameters $R_\mu$, known as rapidities or pair energies, sum to give the energy and are obtained by solving a set of nonlinear Richardson equations.  Two important limiting cases are known.  First, as the interaction strength $G$ tends to infinity, the exact ground state solution becomes number-projected BCS.  Second, for large enough interacting strength $G$, the lowest energy mean-field solution breaks particle number symmetry, and in the thermodynamic limit, this number-broken BCS solution gives the exact energy.\cite{Richardson1977,Roman2002}

Our interest at present is not in the exact solution of this problem.  Rather, we would like to solve the Hamiltonian using more conventional many-body methods, under the assumption that if our methods provide accurate solutions for the pairing Hamiltonian then they should be able to do the same for related but not solvable Hamiltonians.  Unfortunately, it is far from clear how best to treat the problem.

For small coupling $G$, one can certainly use traditional coupled cluster theory.\cite{CoesterKummel,BartlettShavitt}  In that limit, what we will call pair coupled cluster doubles\cite{Dukelsky2003,Ayers2013a,Ayers2013b,Stein2014,Henderson2014b} (pCCD) provides essentially exact results.  In pCCD, we would write the wave function as
\begin{subequations}
\begin{align}
|\Psi_\mathrm{pCCD}\rangle &= \mathrm{e}^{T_2} |0\rangle,
\label{Eqn:pCCD_wfn}
\\
T_2 &= \sum_{ia} t_{ia} \, P_a^\dagger \, P_i,
\label{Eqn:T2CC}
\end{align}
\end{subequations}
where $|0\rangle$ is a mean-field reference occupying the $n_p$ single-particle levels with lowest energies $\epsilon_p$ (\textit{i.e.} a Fermi vaccum) and where indices $i$ and $a$ run over single-particle levels occupied and empty in $|0\rangle$, respectively.  The energy and amplitudes $t_{ia}$ can be readily extracted by solving the Schr\"odinger equation projectively in the subspace of the reference determinant and all double excitations out of it, as
\begin{subequations}
\begin{align}
E &= \langle 0 | \mathrm{e}^{-T_2} \, H \, \mathrm{e}^{T_2} |0\rangle,
\\
0 &= \langle 0 | P_i^\dagger \, P_a \, \mathrm{e}^{-T_2} \, H \, \mathrm{e}^{T_2} |0\rangle.
\end{align}
\end{subequations}

Note that because seniority is a symmetry of the Hamiltonian, excitations which break pairs (\textit{i.e.} those which result in a determinant with singly-occupied orbitals) can be excluded \textit{a priori}; if one were to include them, one would find that they would have zero amplitude.  We have taken advantage of this fact to write $T_2$ in terms only of pair excitations in which both electrons are removed from the same occupied orbital $i$ and placed in the same virtual orbital $a$.  Moreover, single- and triple-excitations necessarily vanish because they must leave at least two singly-occupied orbitals.  Thus, the first correction to pCCD for this Hamiltonian comes from what are called connected quadruple excitations created by $T_4 = \sum_{i>j,a>b} t_{ijab} \, P_a^\dagger \, P_b^\dagger \, P_i \, P_j$.  The exceptional accuracy of pCCD for the weakly correlated case is presumably because true four-body effects are small for weakly correlated systems.  

For repulsive interactions ($G < 0$) pCCD is very accurate even for large $G$.  On the other hand, as $G$ increases in the attractive pairing Hamiltonian, pCCD begins to overcorrelate wildly, and for $G$ not much larger than $G_c$ (where $G_c$ denotes the point at which the Hartree-Fock mean-field reference develops an instability toward a number-broken BCS determinant) the pCCD amplitudes $t_{ia}$ become complex, as does the energy predicted by pCCD.\cite{Dukelsky2003,Henderson2014}  Once this happens, the method is of no real utility.  One could attempt to alleviate this problem by embracing number symmetry breaking and using a BCS coupled cluster approach,\cite{Henderson2014,Signoracci2015,Duguet2015} but while such an approach is reasonable for large enough attractive $G$, the method is in this case not ideal.  The accuracy of the method for $G$ not much larger than $G_c$ is not outstanding, and the BCS coupled cluster produces an artificial first-order phase transition at $G_c$.\cite{Henderson2014}  Moreover, the coupled cluster wave function remains symmetry-broken, so even when the energy is accurate the wave function is unphysical unless we are working in the thermodynamic limit.  Alternatively, one could continue working with a more sophisticated number-preserving theory, and computationally perhaps the simplest way to move beyond pCCD is to work within the pair extended coupled cluster doubles (pECCD) framework,\cite{Henderson2015} in which one approximately includes the effects of $T_4$ in a factorized sort of way by introducing a second set of de-excitation amplitudes.  While pECCD does succeed in salvaging the situation somewhat in that it at least predicts real energies for all $G$, results for large $G$ are poor.

\begin{figure}[t]
\includegraphics[width=\columnwidth]{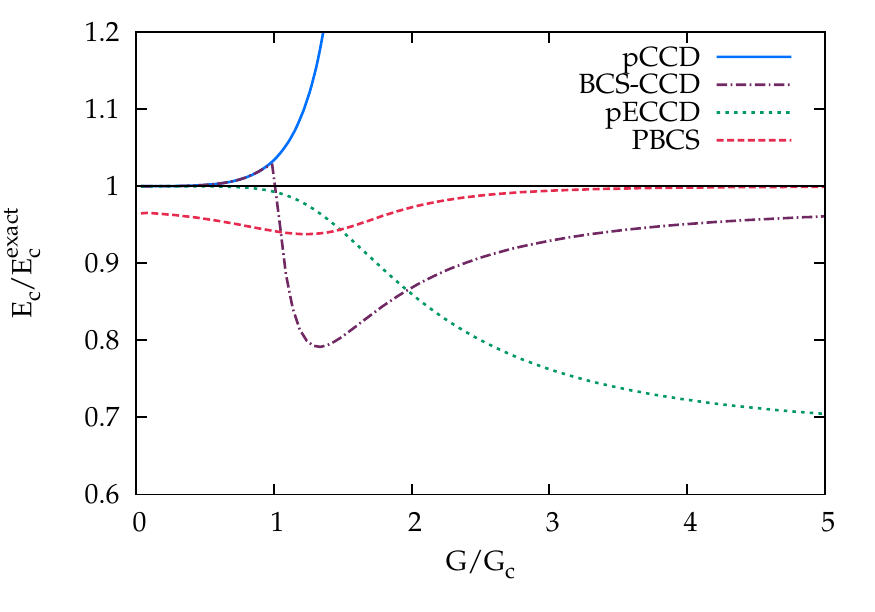}
\caption{Fraction of correlation energy recovered with respect to Hartree-Fock in the 16-site, half-filled pairing Hamiltonian with equally spaced levels.  We have defined $G_c$ as the point at which number symmetry is spontaneously broken by the mean-field.  The pCCD curve is truncated; for larger $G$ the method predicts a complex total energy.
\label{Fig:PriorResults}}
\end{figure}

While pCCD provides the right answer for small $G$ but fails for $G > G_c$, the opposite is true of number-projected BCS (PBCS).  This model provides all the right large $G$ behavior but is not particularly accurate for small $G$.  As $G$ becomes infinite, the single-particle part of the Hamiltonian can be neglected, and it is easy to show that in this case, PBCS delivers an exact eigenstate of the Hamiltonian; as BCS and PBCS give the same energy per particle in the thermodynamic limit (for which $G_c \to 0$), BCS also gives the right answer for large $G$ in this limit.  Figure \ref{Fig:PriorResults} depicts this basic state of affairs for the half-filled pairing Hamiltonian with equally spaced levels ($\epsilon_p = p$).

Ideally, we would like to interpolate between pCCD and PBCS in some way, but it is far from clear how to do so.  The wave functions are expressed in very different ways, to be sure, but the problem is more basic than that: in pCCD the energy and amplitudes are determined by a projective solution to the Schr\"odinger equation, while in PBCS the energy is taken as an expectation value which is variationally minimized with respect to the parameters of the theory.  While pCCD is size extensive (\textit{i.e.} the correlation energy per particle in the thermodynamic limit is non-zero), PBCS is size intensive (the correlation energy in the thermodynamic limit is non-zero, but the correlation energy per particle vanishes).  The two methods do not, it would appear, have much in common.

In this manuscript, we attempt to tackle that challenge, showing how to smoothly go from one model to the other by abandoning both the variational principle needed for PBCS and the exponential form needed for pCCD.  To this end, we will provide more details about the two basic theories in Sec. \ref{Sec:pCCDandPBCS}.  In Sec, \ref{Sec:PoST} we show how to reconcile these two theories \textit{via} more general polynomial similarity transformations and provide results for our new approach.  Section \ref{Sec:Perspective} provides an alternative perspective on our polynomial similarity transformation (PoST) ansatz, and we conclude in Sec. \ref{Sec:Conclusions}.

\section{Pair Coupled Cluster Doubles and Projected BCS Theories
\label{Sec:pCCDandPBCS}}
Let us begin our discussion with examining in more detail how and why coupled cluster doubles fails for the pairing Hamiltonian.  In coupled cluster theory, our aim is to construct a similarity-transformed Hamiltonian
\begin{equation}
\bar{H} = \mathrm{e}^{-T} \, H \, \mathrm{e}^T.
\end{equation}
While $\exp(-T)$ is the inverse of $\exp(T)$, the similarity transformation is non-unitary because $T$ is not anti-Hermitian; accordingly, $\bar{H}$ is non-Hermitian.  Nonetheless, one can choose that similarity transformation so that the right-hand ground-state eigenfunction of $\bar{H}$ is the mean-field reference:
\begin{equation}
\bar{H} |0\rangle = E |0\rangle.
\end{equation}
The amplitude equations of coupled cluster theory facilitate this:
\begin{subequations}
\begin{align}
E &= \langle 0 | \bar{H} |0\rangle,
\\
0 &= \langle \mu | \bar{H} |0\rangle
\end{align}
\end{subequations}
where $\langle \mu|$ stands for any state orthogonal to the reference.  If one uses a full cluster operator
\begin{equation}
T = \sum\limits_{n=1}^{n_p} T_{2n}
\end{equation}
where $T_{2n}$ creates $2n$-fold excitations (recall that odd excitation levels do not contribute because they necessarily break pairs and the Hamiltonian preserves seniority), one can satisfy all of the amplitude equations and in so doing make the reference $|0\rangle$ an exact eigenstate of $\bar{H}$; accordingly, the wave function
\begin{equation}
|\Psi\rangle = \mathrm{e}^T |0\rangle
\label{Eqn:CCwfn}
\end{equation}
is an exact eigenstate of the original Hamiltonian.

\begin{figure}[t]
\includegraphics[width=\columnwidth]{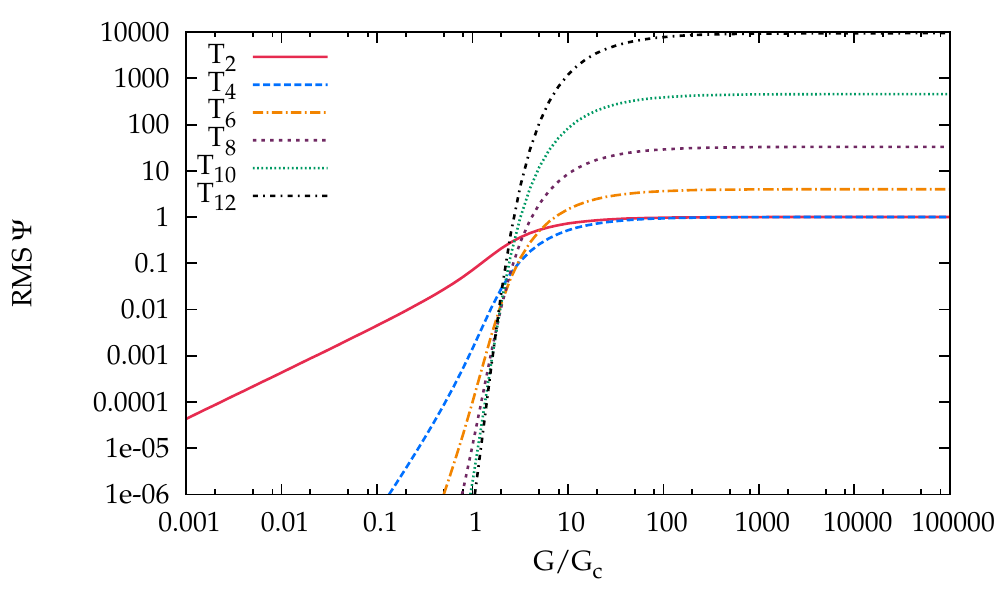}
\caption{Root-mean-square values of amplitudes defining excitation operators in the 12-site, half-filled pairing Hamiltonian with equally spaced levels using the coupled cluster parameterization of the wave function (Eq. \ref{Eqn:CCwfn}).  Results for larger systems are broadly similar.
\label{Fig:RMSCC}}
\end{figure}

In practical calculations, we must of course truncate the cluster operator -- pCCD truncates it, for example, as just $T = T_2$ -- and in this case we cannot simultaneously satisfy all of the amplitude equations.  In pCCD, for example, we satisfy
\begin{equation}
0 = \langle 2 | \mathrm{e}^{-T} \, H \, \mathrm{e}^T |0\rangle
\end{equation}
where $\langle 2|$ stands for the collection of doubly-excited states, but the residuals
\begin{equation}
R_{2n} = \langle 2n | \mathrm{e}^{-T} \, H \, \mathrm{e}^T |0\rangle
\end{equation}
for the $2n$-tuply excited determinants $\langle 2n|$ with $n > 1$ are beyond our control.  When these residuals are small, pCCD will be a good approximation.  Moreover, when the residuals are small at the doubles-only level ($T = T_2$) then the amplitudes defining the higher-order cluster operators $T_4$, $T_6$, and so on will likewise be small.  On the other hand, when the higher-order cluster amplitudes are large or, put differently, the higher-order residuals at the doubles-only level are big, pCCD will fail.  For the pairing Hamiltonian at large $G$, we fall into the latter category, as can be seen in Fig. \ref{Fig:RMSCC}, where we show the root-mean-square size of the amplitudes in various cluster operators as extracted from the exact wave function.

\begin{figure}[t]
\includegraphics[width=\columnwidth]{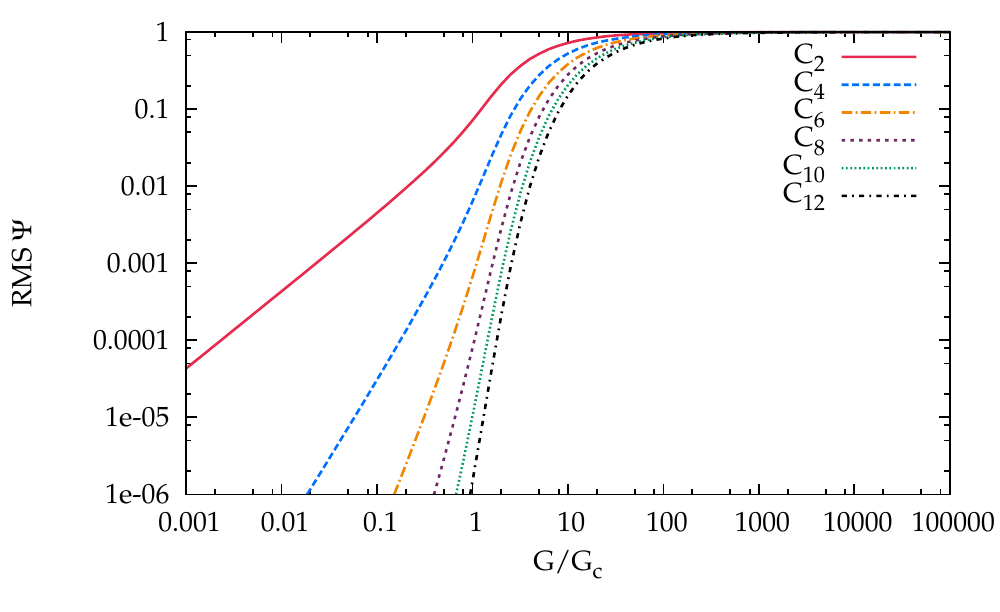}
\caption{Root-mean-square values of amplitudes defining excitation operators in the 12-site, half-filled pairing Hamiltonian with equally spaced levels using the configuration interaction parameterization of the wave function (Eq. \ref{Eqn:CIwfn}).  Results for larger systems are broadly similar.
\label{Fig:RMSCI}}
\end{figure}

What Fig. \ref{Fig:RMSCC} makes clear is that the higher-order cluster operators, far from being negligible, are actually very large.  This is so even though from a configuration interaction (CI) perspective, in which we write
\begin{equation}
|\Psi \rangle = \left(1 + C_2 + C_4 + \ldots\right) |0\rangle
\label{Eqn:CIwfn}
\end{equation}
where $C_{2n}$ creates $2n$-fold excitations, the various coefficients defining the operators $C_{2n}$ are not too large, as seen in Fig. \ref{Fig:RMSCI}.  In other words, the exponential parameterization
\begin{equation}
\mathrm{e}^T = 1 + C
\end{equation}
is in a sense responsible for the failure of pCCD in this case: even when the higher order excitation amplitudes in $C_{2n}$ are not too large, higher order cluster amplitudes in $T_{2n}$ are enormous, so that they cannot be neglected; quite generally, we cannot expect a truncated coupled cluster method to work well for the pairing Hamiltonian.  There is, in other words, no natural truncation of the coupled cluster hierarchy in the strongly correlated limit. The success of pCCD in the weakly correlated limit is rooted in the existence of a natural truncation hierarchy in which $T_2 > T_4 > T_6 > \ldots$ so that the quadruple excitation operator $C_4$ is accurately approximated as $C_4 \approx 1/2 \, T_2^2$, and similarly for higher excitations.

If the conventional exponential framework has no natural truncation for large $G$, the question then becomes whether there is an alternative wave function parameterization in which such a natural truncation takes place.  In other words, is there \textit{any} form in which the wave function can be accurately approximated in terms only of low-order connected excitation operators?  For the reduced BCS Hamiltonian, the solution is apparently provided by projected BCS, a size intensive, seemingly unrelated wave function that one obtains via a simple symmetry projection on a broken symmetry BCS determinant.\cite{PQT,PHF}  Both pCCD and PBCS are geminal theories, but of very different character.  One can write the pCCD wave function of Eqn. \ref{Eqn:pCCD_wfn} in a form which makes it clear that it has $N$ distinct geminals for an $N$-pair system, each spreading over one occupied state and all virtual states:\cite{Ayers2013b}
\begin{subequations}
\begin{align}
|\mathrm{pCCD}\rangle &= \prod_i \Gamma_i^\dagger |-\rangle,
\\
\Gamma_i^\dagger &= P_i^\dagger + \sum_a t_{ia} P_a^\dagger,
\end{align}
\end{subequations}
where $|-\rangle$ is the physical vacuum.  In PBCS, on the other hand, the pairs condense into a single geminal which spreads over every single-particle state:
\begin{subequations}
\begin{align}
|\mathrm{PBCS}\rangle &= \left(\Gamma^\dagger\right)^N |-\rangle,
\\
\Gamma^\dagger &= \sum_i x_i \, P_i^\dagger + \sum_a x_a P_a^\dagger.
\end{align}
\end{subequations}
The question becomes, can we straightforwardly express PBCS in a language that makes it compatible with the coupled cluster ansatz?  If this can be done, then one can look for simple means to blend the two theories.

This can in fact be done. The key is to rewrite the PBCS wave function in terms of particle-hole excitations out of the Fermi vacuum $|0\rangle$.  This, however, is straightforward.  Recall that the BCS wave function itself can be written as a Thouless transformation of the Fermi vacuum, so that one has simply\cite{RingSchuck,Dukelsky2000}
\begin{equation}
|\mathrm{PBCS}\rangle = \mathcal{P}_N \, \mathrm{e}^Z |0\rangle
\end{equation}
where $\mathcal{P}_N$ is the number projection operator and the Thouless transformation is specified by Z:
\begin{subequations}
\begin{align}
Z &= T_1^{(+)} + T_1^{(-)},
\\
T_1^{(+)} &= \sum_a x_a \, P_a^\dagger,
\\
T_1^{(-)} &= \sum_i \frac{1}{x_i} \, P_i,
\end{align}
\end{subequations}
where $a$ and $i$, as before, index single-particle levels empty (occupied) in the Hartree-Fock mean-field reference $|0\rangle$.  The parameters $x_a$ and $x_i$ are given by
\begin{equation}
x_p = \frac{v_p}{u_p}
\end{equation}
where $u_p$ and $v_p$ define the quasiparticle transformation which gives the broken-symmetry mean-field.

\begin{figure}[t]
\includegraphics[width=\columnwidth]{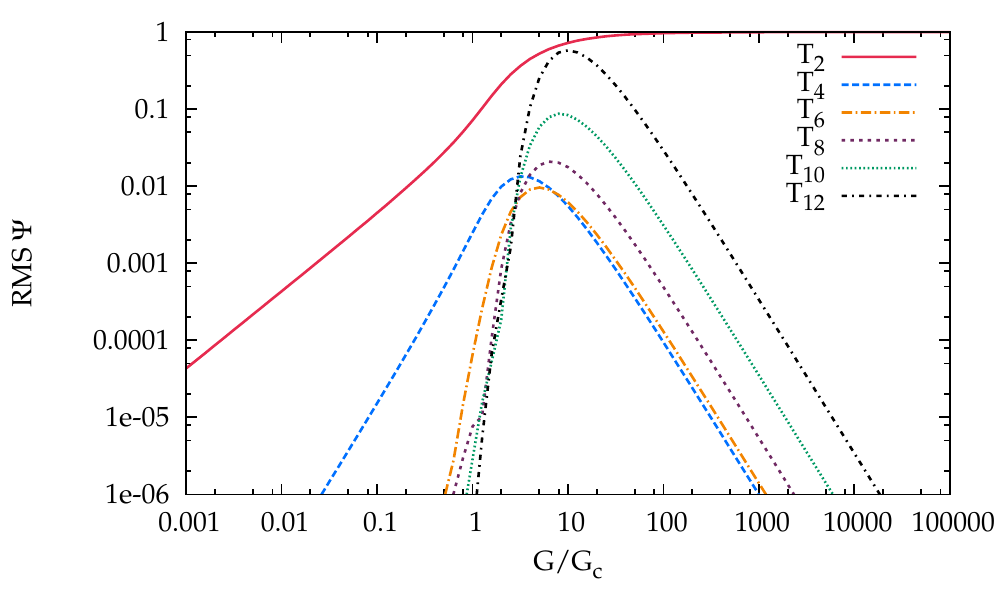}
\caption{Root-mean-square values of amplitudes defining excitation operators in the 12-site, half-filled pairing Hamiltonian with equally spaced levels using the Bessel parameterization of the wave function (Eq. \ref{Eqn:PBCSwfn}).  Compare to the coupled cluster and configuration interaction parameterizations in Figs. \ref{Fig:RMSCC} and \ref{Fig:RMSCI}.  Results for larger systems are broadly similar, although higher excitation amplitudes become fairly large before decaying to zero for large enough $G/G_c$.
\label{Fig:RMSPST}}
\end{figure}

Because $P_a^\dagger$ and $P_i$ commute, we could equivalently write
\begin{subequations}
\begin{align}
|\mathrm{PBCS}\rangle
 &= \mathcal{P}_N \, \mathrm{e}^{T_1^{(+)}} \, \mathrm{e}^{T_1^{(-)}}  |0\rangle
\\
 &= \mathcal{P}_N \, \sum_{mn} \frac{1}{m!} \, \frac{1}{n!} \, \left(T_1^{(+)}\right)^m  \, \left(T_1^{(-)}\right)^n |0\rangle
\\
 &= \sum_{n} \frac{1}{(n!)^2} \left(\sum_{ai} \frac{x_a}{x_i} \, P_a^\dagger \, P_i\right)^n |0\rangle
\end{align}
\end{subequations}
where in the last line we have used that the number projection just picks out the diagonal $m=n$ term in the sum.  We can identify a (factorized) double excitation operator
\begin{equation}
T_2 = \sum_{ai} \frac{x_a}{x_i} \, P_a^\dagger \, P_i
\label{Eqn:T2PBCS}
\end{equation}
in which case we see that the PBCS wave function is just
\begin{equation}
|\mathrm{PBCS}\rangle = \sum \frac{1}{(n!)^2} \, T_2^n |0\rangle = I_0(2 \, \sqrt{T_2})
\end{equation}
where $I_0$ is a modified Bessel function of the first kind.  The accuracy of PBCS suggests that $T_2$ is the only ingredient necessary provided that we are willing to abandon the exponential parameterization of the wave function.  In Fig. \ref{Fig:RMSPST} we show the root-mean-square amplitudes in this Bessel-like parameterization of the wave function
\begin{equation}
|\Psi\rangle = \sum \frac{1}{(n!)^2} \, T^n |0\rangle
\label{Eqn:PBCSwfn}
\end{equation}
or equivalently
\begin{equation}
I_0(2 \, \sqrt{T}) = 1 + C;
\end{equation}
clearly, as $G$ gets larger, the doubles-only approximation becomes very accurate, \textit{i.e.} the higher-order amplitudes vanish as they should.

\begin{figure}[t]
\includegraphics[width=\columnwidth]{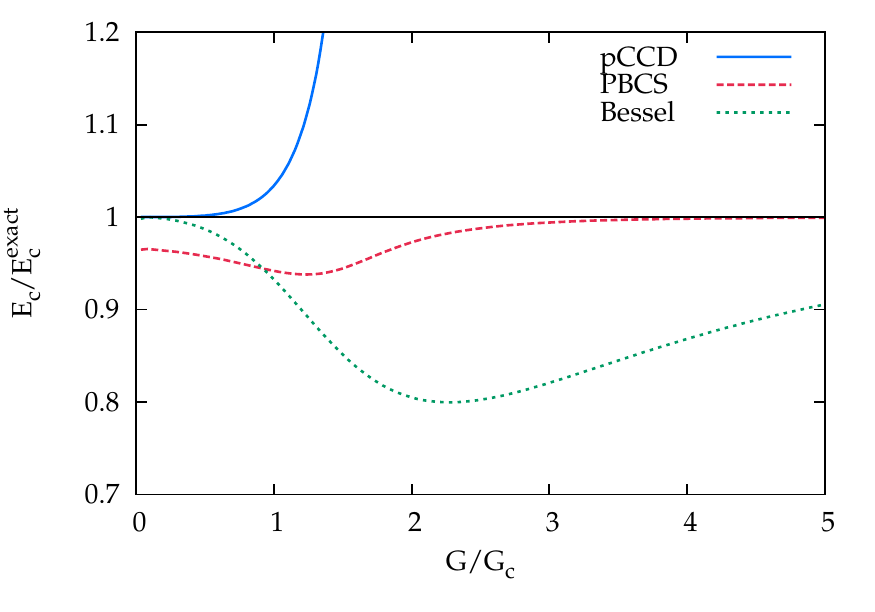}
\caption{Fraction of correlation energy recovered in the half-filled 16-site pairing Hamiltonian as a function of $G$.  We show pCCD and PBCS, while ``Bessel'' denotes the use of our PoST equations (see below) with $c_2 = 1/4 \Leftrightarrow \alpha = 2$.  Put differently, ``Bessel'' indicates a wave function of PBCS form (Eqn. \ref{Eqn:PBCSwfn}) where $T_2$ has been obtained projectively as in pCCD.
\label{Fig:JustifyInterpolation}}
\end{figure}

Because this Bessel function parameterization of the wave function is so well behaved for large $G$, one may be tempted to simply use it everywhere. This temptation should be avoided, at least if one wishes to solve for the wave function in the projective manner outlined in the following section, as can be seen from Fig. \ref{Fig:JustifyInterpolation} where explicitly we solve Eqns. \ref{Eqn:DefHBar} and \ref{Eqn:PoST} with the coefficients defined by the Bessel form ($c_n = \frac{1}{(n!)^2}$).

What we would like, then, is a simple form for the wave function that can interpolate between the exponential form, which works well for small $G$, and the Bessel form, which works well for large $G$.  A natural parameterization is to write simply
\begin{subequations}
\begin{align}
|\Psi\rangle &= F_\alpha(T) |0\rangle,
\\
F_\alpha(T) &= \sum \frac{1}{(n!)^\alpha} \, T^n
\label{Eqn:DefFT}
\end{align}
\end{subequations}
where as $\alpha$ goes from 1 to 2, $|\Psi\rangle$ goes from the pCCD to the PBCS form.  The remainder of this manuscript examines the implications of such a choice, when combined with similarity transformations which require the construction of $F^{-1}(T)$.

\section{Polynomial Similarity Transformations
\label{Sec:PoST}}
Let us begin with a more general discussion of polynomial similarity transformation, which we will specialize to the wave operator (the operator which maps the reference state $|0\rangle$ onto the desired state $|\Psi\rangle$) defined in Eqn. \ref{Eqn:DefFT} presently.

Suppose, then, that we have a wave operator parameterized in terms of an excitation operator $T$ as
\begin{equation}
F(T) = 1 + T + c_2 \, T^2 + c_3 \, T^3 + \ldots
\label{DefFt}
\end{equation}
where we can always choose the coefficient of $T^0$ to be one because this is equivalent to choosing the wave function
\begin{equation}
|\Psi_F \rangle = F(T) |0\rangle
\end{equation}
to have intermediate normalization ($\langle 0 | \Psi_F \rangle = 1$), and we can similarly choose the coefficient of $T^1$ to be one since any other choice can be absorbed into the definition of $T$.  Because $T$ is a pure excitation operator, there is some $n$ beyond which $T^n = 0$, so, provided that $T$ is finite, we can always define an inverse operator
\begin{equation}
F^{-1}(T) = 1 - T - (c_2 - 1) \, T^2 - (c_3 - 2 \, c_2 + 1) \, T^3 + \ldots
\end{equation}
such that $F^{-1}(T) \, F(T) = 1$.  We can accordingly define a similarity-transformed Hamiltonian
\begin{subequations}
\begin{align}
\bar{H} &= F^{-1}(T) \, H \, F(T)
\\
        &= H + [H,T] + c_2 \, [[H,T],T]
\\
        &+ \left(2 \, c_2 - 1\right) T \, [H,T] + \mathcal{O}(H \, T^3)
\nonumber
\end{align}
\label{Eqn:DefHBar}
\end{subequations}
whose eigenvalues are the same as the eigenvalues of the original Hamiltonian $H$.  When $c_n = 1/n!$ the similarity-transformed Hamiltonian is purely connected (\textit{i.e.} expressible entirely in terms of nested commutators of $H$ with $T$), but in general it is not.  Note that the term with coefficient $\left(2 \, c_2 - 1\right)$ is disconnected and does not appear in traditional coupled cluster theory.

We will be interested here in the doubles-only theory ($T = T_2$) in which, as with coupled-cluster theory, we will insist that the Fermi vacuum is an approximate right-hand eigenstate of $\bar{H}$:
\begin{subequations}
\begin{align}
E &= \langle 0 | \bar{H} | 0 \rangle,
\\
0 &= \langle 2 | \bar{H} | 0 \rangle.
\end{align}
\label{Eqn:PoST}
\end{subequations}
In this case, we can disregard all terms of order $H \, T^3$ or higher as they cannot contribute to the amplitude equations.  In practice, the foregoing equations could be implemented in a traditional coupled-cluster doubles code by making two changes: every $T^2$ term in the amplitude equations must be multiplied by $2 \, c_2$, and to the orbital energy denominator one must add $(1 - 2 \, c_2) \, E_c$, where $E_c = \langle 0 | H \, T |0\rangle$ is the correlation energy.  Alternatively, one could leave the denominator unchanged and add $(2 \, c_2 - 1) \, E_c \, T_2$ to the numerator.  This last term arises from the projection of the disconnected term:
\begin{equation}
\langle 2 | T \, [H,T] |0\rangle = \langle 2 | T | 0\rangle \langle 0 | H \, T |0\rangle.
\end{equation}
This closed disconnected term in the amplitude equations is what is known as unlinked and is an inevitable consequence of demanding that we solve the similarity-transformed Schr\"odinger equation with a non-exponential similarity transformation.  We shall say more about this later.

As we have noted, we will make the specific choice $c_2 = 1/2^\alpha$ where for $\alpha=1 \Longrightarrow c_2 = 1/2$ we reduce to coupled cluster doubles (or, for the pairing Hamiltonian, pCCD).  For any choice of $\alpha$ the method will be exact for two-particle systems (because for any choice of $\alpha$ we are carrying out an exact similarity transformation and solving the similarity-transformed Schr\"odinger equation exactly, presuming that single excitations have been eliminated by a careful choice of the Fermi vacuum).  Only for $\alpha = 1$ is the method extensive because only for $\alpha = 1$ does the unlinked term disappear.  This means that while the theory is exact for any two-electron system, it is in general \textit{not} exact for a collection of non-interacting two-electron units.

\begin{figure}[t]
\includegraphics[width=\columnwidth]{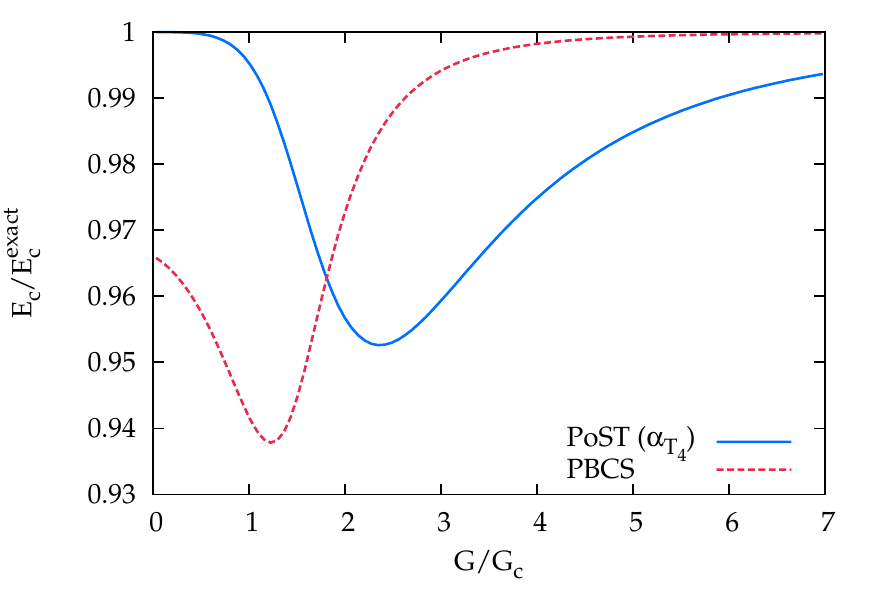}
\caption{Fraction of correlation energy recovered in the half-filled 16-site pairing Hamiltonian as a function of $G$.  We show PBCS and our PoST method choosing that parameter $\alpha$ which minimizes $T_4$ given the exact $T_2$.
\label{Fig:EFromAlphaT4}}
\end{figure}

\begin{figure}[t]
\includegraphics[width=\columnwidth]{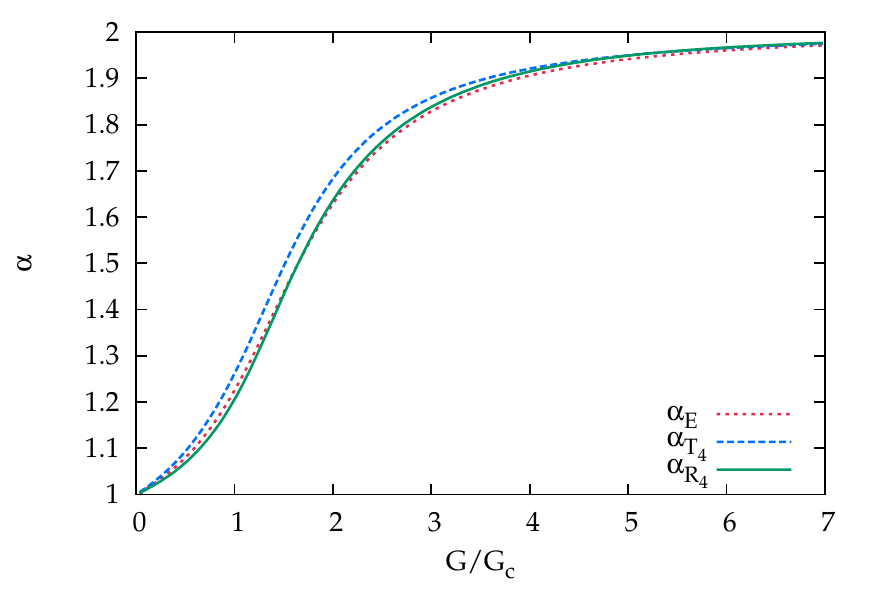}
\caption{The $\alpha$ parameter defining the PoST correlator $F(T)$ as a function of $G$ in the half-filled 16-site pairing Hamiltonian.  We include the $\alpha$ that minimizes $T_4$ given the exact $T_2$ (labeled as $\alpha_{T_4}$) and the $\alpha$ which gives the exact energy (labeled as $\alpha_E$).  We also include the $\alpha$ that minimizes the $R_4$ residual (see below), labeled as $\alpha_{R_4}$.  Note that in all cases $\alpha$ goes from 1 at small $G$ (as expected from pCCD) to 2 at large $G$ (as expected from PBCS).  
\label{Fig:CompareAlphas}}
\end{figure}

The question becomes how such a general polynomial similarity transformation (PoST) performs for the pairing Hamiltonian.  The results will clearly depend crucially on the correct choice of $\alpha$.  We posit that the correct $\alpha$ should make higher excitations negligible.  Thus, for example, if one were to write
\begin{equation}
C_4 = T_4 + \frac{1}{2^\alpha} \, T_2^2
\end{equation}
where $C_4$ is the quadruple excitation operator from configuration interaction, we wish to find the value of $\alpha$ which makes $T_4$ minimal.  Figure \ref{Fig:EFromAlphaT4} shows the results of this choice.  Specifically, we have first solved for the exact $C_4$ and $T_2 = C_2$ and chosen the value of $\alpha$ minimizing $T_4$ defined above; we have then used this value of $\alpha$ in the doubles-only theory outlined in Eqns. \ref{Eqn:PoST} and \ref{Eqn:DefHBar} with $c_2 = 1/2^\alpha$.  We see that this choice of $\alpha$ yields highly accurate energies.  In fact, the value of $\alpha$ so chosen is fairly similar to the value of $\alpha$ chosen to yield the exact energy, as seen in Fig. \ref{Fig:CompareAlphas}.

Of course in practice any method which requires us to first solve the problem exactly so that we can extract parameters with which to solve the problem approximately is not useful.  The foregoing results, however, suggest a way forward: one should choose the parameter $\alpha$ so as to minimize the quadruple-excitation residual $R_4 = \langle 4 | \bar{H} |0\rangle$ in our doubles-only theory.  In the exact theory, $R_4$ vanishes and is linear in $T_4$; thus, minimizing $R_4$ in the absence of $T_4$ suggests that the $T_4$ needed to eliminate this residual should be small.  This, therefore, completes our specification of the theory henceforth:
\begin{subequations}
\begin{align}
F(T_2) &= \sum \frac{1}{(n!)^\alpha} \, T_2^n,
\\
\bar{H} &= F^{-1}(T_2) \, H \, F(T_2),
\\
E &= \langle 0 | \bar{H} | 0\rangle,
\\
0 &= \langle 2 | \bar{H} |0 \rangle,
\\
\alpha &= \mathrm{arg \; min} \sum_4 |\langle 4 | \bar{H} |0 \rangle|^2.
\end{align}
\label{Eqn:PoSTComplete}
\end{subequations}
In the preceding equation, the summation over ``4'' is meant to indicate summation over quadruply-excited determinants $\langle 4|$.  We note in passing that other alternatives for the determination of $\alpha$ have been explored, but as minimizing the $R_4$ residual is computationally and conceptually simple and delivers excellent results, we do not consider such alternatives here.  We also note that this choice makes the method generally exact in the pairing Hamiltonian with two pairs in four levels, because the exact wave function in this case can be written as
\begin{equation}
|\Psi\rangle = \left(1 + T_2 + \frac{1}{2} \, T_2^2 + T_4\right) |0\rangle
\end{equation}
and there is a single coefficient in $T_4$, \textit{i.e.} we can parameterize $T_4 = \lambda T_2^2$.  Provided, then, that $\left(1/2 + \lambda\right)$ can be parameterized as $1/2^\alpha$, we can simultaneously make both the $R_2$ and $R_4$ residuals vanish, which is equivalent to exactly solving the Schr\"odinger equation for four particles when single and triple excitations vanish.

In order to facilitate correct derivation and efficient implementation, the expressions for the $R_4$ residuals and their derivatives with respect to $\alpha$ were obtained from an algebraic generator and implemented by an automatic code generator.  Details about these tools will be presented in due time.\cite{Jinmo}  For the version of the theory outlined above with $T_2$ taking the pCCD form, the computational scaling is $\mathcal{O}\left(o^2 \, v^2\right)$, where pCCD itself has $\mathcal{O}\left(o \, v^2\right)$ scaling.

\begin{figure}[t]
\includegraphics[width=\columnwidth]{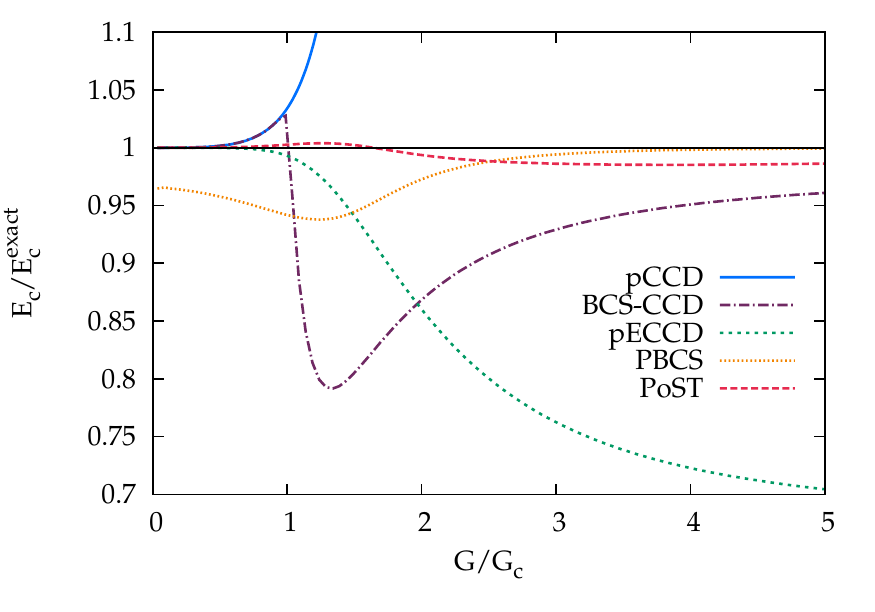}
\caption{Fraction of correlation energy recovered in the half-filled 16-site pairing Hamiltonian as a function of $G$.
\label{Fig:ProjectedEnergy}}
\end{figure}

\begin{figure}[t]
\includegraphics[width=\columnwidth]{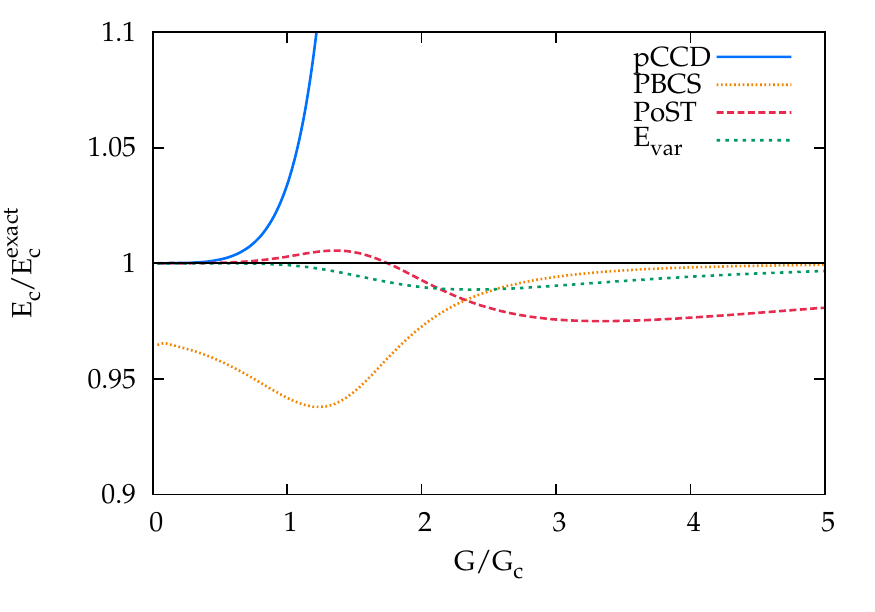}
\caption{Fraction of correlation energy recovered in the half-filled 16-site pairing Hamiltonian as a function of $G$.  We show pCCD and PBCS, along with our new PoST model and the expectation value of the PoST wave function as defined in Eqn. \ref{Eqn:ExpVal}.
\label{Fig:ExpVal}}
\end{figure}

In Fig. \ref{Fig:ProjectedEnergy} we show that this polynomial similarity transformation is exceptionally accurate for the attractive pairing Hamiltonian across various values of $G$.  The corresponding $\alpha$ is shown in Fig. \ref{Fig:CompareAlphas}.  Like pCCD and unlike PBCS, the PoST energy expression is projective and thus not variationally bound, but we can take the Hermitian expectation value after obtaining the amplitudes projectively and evaluate
\begin{equation}
E_\mathrm{var} = \frac{\langle 0| F^\dagger(T) \, H \, F(T) |0 \rangle}{\langle 0 | F^\dagger(T) \, F(T) |0\rangle}
\label{Eqn:ExpVal}
\end{equation}
which compares more readily with PBCS.  We show the expectation value form of the energy in Fig. \ref{Fig:ExpVal} to illustrate how well the expectation value reproduces the PBCS energy.  Note, however, that evaluation of the expectation value has combinatorial cost and is therefore not a practical approach in general.

\begin{figure}[t]
\includegraphics[width=\columnwidth]{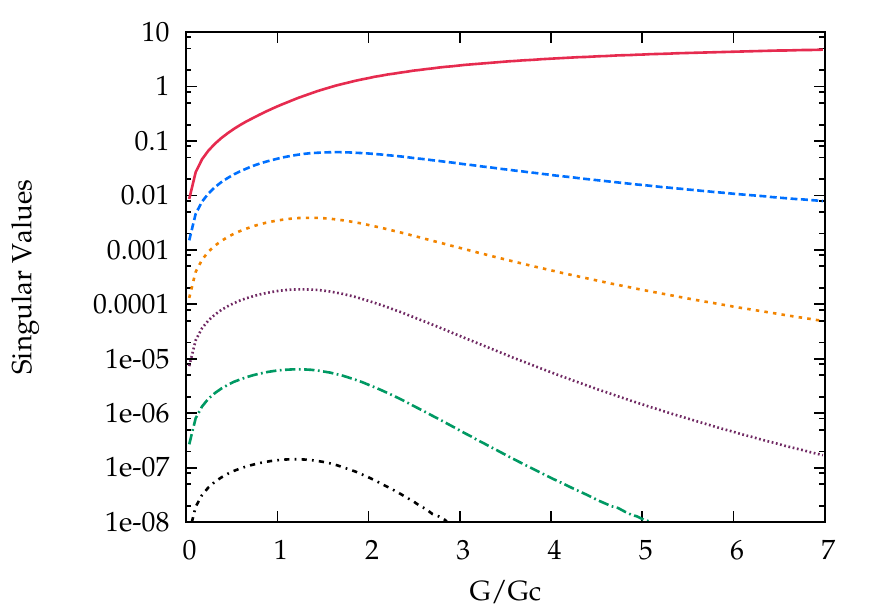}
\caption{Singular values of the PoST $T_2$ amplitudes in the half-filled 16-site pairing Hamiltonian as a function of $G$.  Note that as $G$ becomes large, all singular values but one tend to zero; in other words, the $T_2$ amplitudes factorize in the PBCS form as $G \to \infty$.
\label{Fig:SVD}}
\end{figure}

In addition to using an expectation value rather than a projective energy expression, PBCS has a second difference from PoST with $\alpha = 2$: in the former, we have a factorizable $T_2$, as we saw in Eqn. \ref{Eqn:T2PBCS}.  It may be interesting to check to what extent $T_2$ calculated from PoST factorizes similarly.  To do so, we perform a singular value decomposition (SVD) of the matrix of amplitudes $t_{ia}$ defining our general $T_2$.\cite{Kinoshita2003}  The SVD permits us to write
\begin{equation}
t_{ia} = \sum_{\mu\nu} y_{i\mu} \, w_{\mu\nu} \, x_{\nu a}
\end{equation}
where the square matrices $\mathbf{x}$ and $\mathbf{y}$ are unitary and where the possibly rectangular matrix $\mathbf{w}$ has non-zero entries (the singular values of $\mathbf{t}$) only on the diagonal ($\mu = \nu$).  If there is only one non-zero singular value of $\mathbf{t}$, then $\mathbf{t}$ factorizes as
\begin{equation}
t_{ia} = y_i \, x_a \, w
\end{equation}
and $w$ can be absorbed into the definitions of $x$ and $y$ to produce amplitudes $t_{ia}$ of the PBCS form.  In Fig. \ref{Fig:SVD} we show the singular values of the PoST amplitudes $\mathbf{t}$ as a function of $G$.  For large $G$, we have only one non-zero singular value, which means that the wave function takes the PBCS form.  

\begin{figure}[t]
\includegraphics[width=\columnwidth]{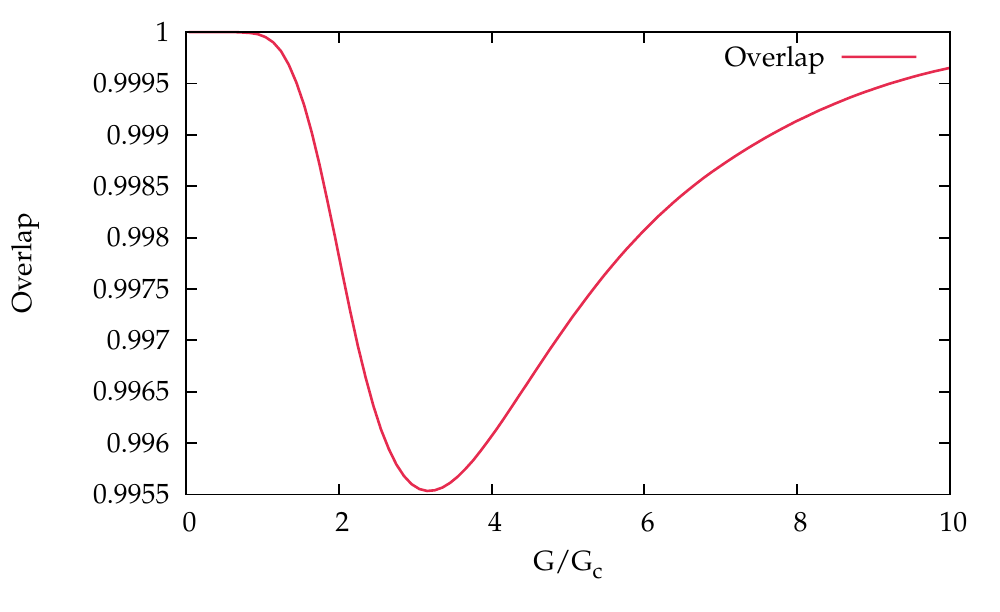}
\caption{Overlap of the PoST wave function $F_\alpha(T) |0\rangle$ with the exact wave function for the half-filled 16-site pairing Hamiltonian as a function of $G$.
\label{Fig:Overlap}}
\end{figure}

While the energetic performance of our PoST approach is highly satisfactory, energies alone are not the only quantity of interest.  We also wish to get, if at all possible, the correct wave function.  The expectation value plotted in Fig. \ref{Fig:ExpVal} suggests that the PoST ket $F(T) |0\rangle$ is very accurate.  This is confirmed by the overlap with the exact wave function, plotted in Fig. \ref{Fig:Overlap}.  The fact that PoST delivers nearly the exact wave function means that it should also deliver essentially exact properties if properties were evaluated as expectation values.  As we have noted previously, this is too cumbersome an approach in practice, and we would prefer to use the response formalism\cite{LinResponse,LinearResponseGus,BartlettShavitt} of standard coupled cluster theory, which we have thus far not tested in combination with PoST.

\section{Perspective
\label{Sec:Perspective}}
Thus far, we have rationalized the polynomial similarity transformations in an essentially \textit{post hoc} manner.  In fact, however, there is a deeper explanation at play which may be instructive to consider here.

Let us, therefore, consider the large $G$ limit of the pairing Hamiltonian.  In the limit where $G$ is very large compared to the single particle level spacing, the Hamiltonian goes to
\begin{equation}
H \to \bar{\epsilon} \, N - G \, P^\dagger \, P
\end{equation}
where $N$ is the total number operator ($N = \sum_p N_p$) and similarly $P$ and $P^\dagger$ are global pairing operators $(P = \sum_p P_p)$, while $\bar{\epsilon}$ is an average single-particle energy level.  The ground state eigenfunction of this Hamiltonian is
\begin{equation}
|\Psi\rangle = \left(P^\dagger\right)^{n_p} |-\rangle
\end{equation}
where recall that $n_p$ is the number of occupied pairs and $|-\rangle$ is the physical vacuum.  This wave function is of PBCS form.  In the language of configuration interaction, all the CI coefficients are identical (and equal to one in intermediate normalization).  Similarly, various coupled cluster amplitudes for various excitation levels are identical.  Generically, we have
\begin{subequations}
\begin{align}
T_2
 &= \sum_{ai} P_a^\dagger \, P_i,
\\
T_{2n}
 &= t_{2n} \, \sum_{a_1 > a_2 > \ldots a_n} \, \sum_{i_1 > i_2 > \ldots i_n} P_{a_1}^\dagger \ldots P_{a_n}^\dagger P_{i_n} \ldots P_{i_1}
\\
 &= \frac{1}{(n!)^2} \, t_{2n} \, T_2^n
\nonumber
\end{align}
\end{subequations}
where we have used the fact that $C_2 = T_2$ to note that the coefficient of all double excitations is one, and in going from the second line to the third have noted that we can relax the summation restrictions at the cost of a factor of $1/(n!)^2$ since there are $n!$ equivalent orders of virtual pair creation operators and $n!$ equivalent orders of occupied pair annihilation operators.

We can use the fact that the ground state wave function is of PBCS form,
\begin{equation}
|\Psi\rangle = I_0(2 \sqrt{T_2}) |0\rangle,
\end{equation}
to extract the coefficients $t_4$, $t_6$, and so on defining higher-order excitation operators in the coupled cluster framework by equating
\begin{subequations}
\begin{align}
T_2 + \frac{1}{4} \, t_4 \, T_2^2 + \frac{1}{36} \, t_6 \, T_2^3 + \ldots
 &= \ln(I_0(2 \, \sqrt{T_2}))
\\
 &= T_2 - \frac{1}{4} \, T_2^2 + \frac{1}{9} \, T_2^3 + \ldots
\end{align}
\end{subequations}
so that $t_4 = -1$, $t_6 = 4$, and so on -- these are the limiting values of the amplitudes plotted in Fig. \ref{Fig:RMSCC}, and are independent of the number of pairs or number of levels.  In other words, analytically the higher-order $T$ amplitudes in a coupled cluster parameterization of the wave function all factor as polynomials of $T_2$ for large $G$, and this factorization is precisely such the wave operator $\exp(T)$ can be resummed as $I_0(2 \, \sqrt{T_2})$ in terms of $T_2$ alone.

The exact $T_2$ equation in a coupled cluster framework is the one that contains all possible contributions from higher cluster operators:
\begin{equation}
0 = \langle 2| H + H \, T_2 - T_2 \, H + \frac{1}{2} \, H \, T_2^2 - T_2 \, H \, T_2 + H \, T_4 |0\rangle
\label{Eqn:ExactT2}
\end{equation}
when $T_1 = T_3 = 0$, as we have in the pairing Hamiltonian.  Inserting $T_4 = -1/4 \, T_2^2$ as we have in the large $G$ limit of the pairing Hamiltonian, we have
\begin{equation}
0 = \langle 2| H + H \, T_2 - T_2 \, H + \frac{1}{4} \, H \, T_2^2 - T_2 \, H \, T_2|0\rangle
\end{equation}
which is precisely our PoST amplitude equation with $\alpha = 2$.  Thus, the PoST amplitude equation in the large $G$ limit delivers the \textit{exact} $T_2$ amplitudes and therefore the exact energy.  Moreover, for large $G$ all higher-order residuals ($R_4$, $R_6$, and so on) vanish because the wave function $|\Psi\rangle = F(T_2) |0\rangle$ becomes an eigenfunction of the Hamiltonian.  For finite $G$, one can view PoST as simply making the ansatz
\begin{equation}
\frac{1}{2} \, T_2^2 + T_4 \approx \frac{1}{2^\alpha} \, T_2^2 \Longrightarrow T_4 \approx \left(\frac{1}{2^\alpha} - \frac{1}{2}\right) \, T_2^2.
\end{equation}
Apparently this ansatz is reasonably accurate for the pairing Hamiltonian with the appropriate choice of $\alpha$ for any value of $G$.

\begin{figure}[t]
\includegraphics[width=\columnwidth]{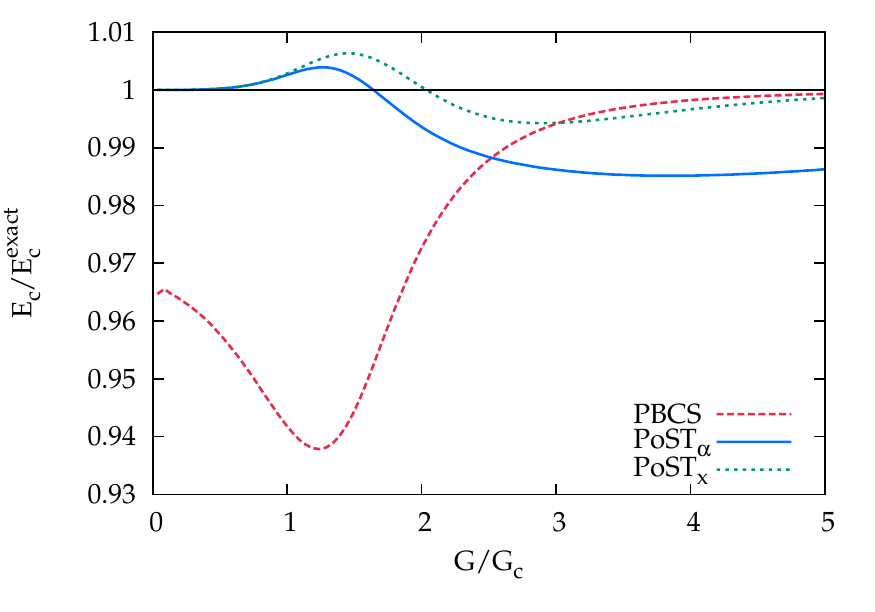}
\caption{Fraction of correlation energy recovered in the half-filled 16-site pairing Hamiltonian as a function of $G$.  We show PBCS, and our PoST model using the similarity transforms defined by $F_\alpha(T)$ given in Eqn. \ref{Eqn:DefFT} and by $F_x(T)$ given in Eqn. \ref{Eqn:DefFxT}, which we label as PoST$_\alpha$ and PoST$_\mathrm{x}$, respectively.
\label{Fig:xModel}}
\end{figure}

These considerations have led us to consider alternative interpolations between pCCD and PBCS.  Without going into great detail here, we could define very simply
\begin{equation}
F_x(T_2) = \mathrm{e}^{(1-x) \, T_2} \, I_0(2 \, \sqrt{x \, T_2})
\label{Eqn:DefFxT}
\end{equation}
where $x$ is an interpolation parameter.  Clearly, as $x$ goes from $0$ to $1$, $F_x(T_2)$ goes between the exponential form defining pCCD and the Bessel form obtained from PBCS.  One could use this interpolation in the scheme summarized in Eqn. \ref{Eqn:PoSTComplete}, merely replacing $F_\alpha(T_2)$ with $F_x(T_2)$ but otherwise following the same general path of using $F_x(T_2)$ to construct a similarity-transformed Hamiltonian $\bar{H}$, solving for the energy and for $T_2$ by left-projection, and obtaining the interpolation parameter $x$ by minimizing the $R_4$ residual.  Preliminary results for this approach are presented in Fig. \ref{Fig:xModel} and are even better than those obtained from $F_\alpha(T)$.

\section{Conclusions
\label{Sec:Conclusions}}
The existence and importance of strong correlations poses somewhat of a conundrum for computational methods.  Weakly correlated systems can be straightforwardly described by essentially standard, black-box techniques.  Strong correlations, however, seem to require specially tailored theoretical approaches which require the user to have a degree of insight into the problem under investigation.  Moreover, these techniques are not, generally, well suited to the description of weak correlations in systems of any reasonable size, if for no other reason than for their computational expense.  Approaches which can simultaneously describe both strong and weak correlations without much fine-tuning on the part of the user are in short supply.

For the reduced BCS Hamiltonian, the proposed PoST method seems likely to fit the bill, forming as it does an interpolation between one theory (pCCD) which accurately describes the weakly-correlated limit and another (PBCS) which accurately describes the strongly-correlated limit.  Provided that one can indeed interpolate between these two limits and provided that one can readily compute a useful interpolation parameter, PoST constitutes a useful theoretical tool for all interaction strengths.  Of course we have a simple alternative for the reduced BCS Hamiltonian -- namely, we could solve the problem exactly instead -- but techniques such a PosT should prove valuable for more general (and more realistic) attractive pairing interactions of the form $-\sum V_{pq} P_p^\dagger P_q$ where no exact solution is available.  Moreover, the techniques presented here can be straightforwardly generated to general two-body Hamiltonians simply by relaxing the restriction that cluster operator $T_2$ take the paired excitation form of Eqn. \ref{Eqn:T2CC}; this is equivalent to generalizing the ansatz in the strongly correlated limit from projected BCS to projected Hartree-Fock-Bogoliubov.

While conceptually it may be simplest to think about PoST as interpolating between two limits, it may be more fruitful to view it as an attempt to approximate the exact $T_2$ equation (Eqn. \ref{Eqn:ExactT2}) by providing a simple form for the missing quadruple excitation coefficients defining $T_4$; in this case, choosing $T_4$ to factor as $\lambda T_2^2$, a factorization which becomes exact in the strongly correlated limit.  In this, our PoST model resembles previous efforts to generalize traditional coupled cluster theory for the description of strong correlations\cite{Paldus1984,Piecuch1996} which focused on projected collinear spin states.  To the best of our knowledge, this is the first such generalization explicitly designed to describe the strong pairing correlations needed for the description of superconductivity.

Let us emphasize that the PoST approach is rather heterodox when viewed from the coupled cluster perspective, because the theory has embraced disconnected terms (terms not expressible solely as commutators) which traditional single-reference coupled cluster theory excludes.  Worse, these disconnected terms in the effective Hamiltonian $\bar{H}$ give rise to unlinked terms in the amplitude equations, yet there are no unlinked terms in the exact theory.\cite{LinkedDiagramTheorem}  A consequence of these unlinked terms is that PoST is not properly extensive (the energy does not scale correctly with system size, though we note in passing that extensivity is a somewhat troublesome concept for the pairing Hamiltonian since, due to the infinite range of its two-body interaction, the exact energy is \textit{quadratic} in particle number).  While the lack of extensivity is unpleasant, we do not regard it as a fatal flaw because we contend that in approximate methods, satisfaction of exact constraints is less important than obtaining accurate results.  Thus, for example, we are happy to use coupled cluster theory even though it is not (as it should be) variational, because coupled cluster theory has been spectacularly successful for a wide variety of problems.  

As we have seen, the attractive pairing Hamiltonian is not one of these problems, essentially because the exponential ansatz of coupled cluster theory is poorly adapted for describing the kinds of strong pairing correlations we see in this case.  The correct physics for large $G$ is instead given by PBCS, which can be written as a double-excitation only theory; the price one pays is the introduction of terms which from a coupled cluster perspective are unnatural.  One could, of course, describe attractive pairing interactions exactly without these unlinked pieces, but this evidently requires the inclusion of very high excitation levels and possibly even all excitations, the cost of which is prohibitive.  In this case, at least, we deem the price of unlinked pieces to be worth paying, allowing us as they do to obtain an exceptionally accurate description of the reduced BCS Hamiltonian at a reasonable computational cost.

We should resolve an apparent inconsistency: there are no unlinked terms in the exact theory, yet we have unlinked terms in our wave function even though that wave function is essentially exact for large $G$.  The key is to note that for large $G$, $T_2$ begins to factorize.  When it factorizes exactly, the unlinked term $E_c T_2$ in the amplitude equation cancels against other quadratic terms, and the final result is linked.  As $G$ tends to infinity, in other words, the theory becomes effectively linked, in a manner analogous to what is seen in collinear spin projection.\cite{Piecuch1996}

Finally, we note while it is not simply traditional coupled cluster theory, PosT is also not far from it.  Thus, the wide variety of tools developed for the latter can be straightfowardly extended to the former.  For example, properties could be evaluated using the response formalism,\cite{LinResponse,LinearResponseGus,BartlettShavitt} and excited states could be accessed through the equation-of-motion methodology.\cite{EOMCC,EOMCC2}  The work we have shown here restricts the double-excitation amplitudes to have zero seniority, but that restriction can be relaxed in general; if one relaxes that restriction, it may or may not be desirable to work with a singlet-paired version\cite{CCD0} of the theory.  While we have presented PoST as a theory for ground state correlations, much more is possible.

\begin{acknowledgments}
This work was supported by the U.S. Department of Energy, Office of Basic Energy Sciences, Computational and Theoretical Chemistry Program under Award No.DE-FG02-09ER16053. G.E.S. is a Welch Foundation Chair (C-0036).  JD acknowleges support from the Spanish Ministry of Economy and Competitiveness through Grant FIS2012-34479.
\end{acknowledgments}

\bibliography{PoST}

\end{document}